\documentclass[aps,prd,10pt,superscriptaddress,nofootinbib,showkeys,twocolumn]{revtex4-2}

\usepackage[T1]{fontenc}
\usepackage{lmodern}
\usepackage{erewhon}
\usepackage[scaled=1.05]{courierten}
\usepackage[utopia]{newtxmath}
\usepackage[bb=dsserif,bbscaled=1.15,scr=rsfs]{mathalpha}

\usepackage{mathtools,siunitx,physics}
\usepackage{graphicx}
\usepackage[normalem]{ulem}
\usepackage{enumerate}
\usepackage{silence}
\ExplSyntaxOn
\msg_redirect_name:nnn{siunitx}{physics-pkg}{none}
\ExplSyntaxOff
\usepackage[dvipsnames]{xcolor}
\definecolor{theme}{rgb}{0.7, 0, 0.35}
\RequirePackage{hyperref}
\hypersetup{colorlinks, hypertexnames=false, allcolors=theme, bookmarksnumbered=true}%
\usepackage[capitalise]{cleveref} % only after hyperref
\usepackage{orcidlink}
\usepackage{xspace}
\usepackage{etoolbox}

\newcommand{\ie}{i.e.,\@\xspace}

\crefname{section}{Sec.}{Secs.}
\Crefname{section}{Section}{Sections}
\crefname{appendix}{App.}{Apps.}
\Crefname{appendix}{Appendix}{Appendices}
\crefname{paragraph}{paragraph}{paragraphs}
\Crefname{paragraph}{Paragraph}{Paragraphs}

\makeatletter
\patchcmd{\subequations}{\crefalias{equation}{subequation}}{\crefalias{equation}{eqs}}{}{\errmessage{subequations patch failed}}
\apptocmd{\subequations}{\crefalias{equation}{\ifnum\c@equation=\z@ eqs\else subequation\fi}}{}{\errmessage{subequations append failed}}
\pretocmd{\appendix}{\crefalias{section}{appendix}\crefalias{subsection}{subappendix}}{}{\errmessage{appendix patch failed}}
\makeatother

\newcommand{\Cite}[1]{Ref.~\cite{#1}}
\newcommand{\Cites}[1]{Refs.~\cite{#1}}

\DeclareMathAlphabet{\mathpzc}{OT1}{pzc}{m}{it}

\newcommand{\mc}[1]{\mathcal{#1}}

\newcommand{\mk}{\mspace{1mu}}
\renewcommand{\vec}[1]{\boldsymbol{\mathrm{#1}}}
\renewcommand{\vu}[1]{\hat{\vec{#1}}}
\newcommand{\pd}{\partial}

\newcommand{\del}{\vec{\nabla}}

\newcommand{\sub}[1]{_\mathrm{#1}}

\newcommand{\bb}{\vec{B}}

\graphicspath{{./}{figs/}{figures/}}

\newcommand{\rs}{r_*}
\newcommand{\rth}{r\sub{mag}}
\newcommand{\Bmin}{B\sub{min}}
\newcommand{\freq}{f\sub{req}}
\newcommand{\nG}{n\sub{G}}
\newcommand{\etareq}{\eta\sub{req}}
\newcommand{\mpckms}{\si{Mpc.km.s^{-1}}}
\newcommand{\ds}{\Delta s}

\begin{document}

\title{
How far do galactic magnetic fields diffuse into cosmic voids?
}

\author{Deepen~Lark\,\orcidlink{0000-0001-5226-1913}}
\email{deepenlark@uni-bonn.de}
\affiliation{Argelander-Institut für Astronomie, Universität Bonn, Bonn 53121, Germany}

\author{Ruth~Durrer\,\orcidlink{0000-0001-9833-2086}}
\email{ruth.durrer@unige.ch}
\affiliation{D\'{e}partement de Physique Th\'{e}orique and Center for Astroparticle Physics, Universit\'{e} de Gen\`{e}ve,  24 quai Ernest Ansermet, 1211 Gen\`{e}ve 4, Switzerland}

\author{Jennifer~Schober\,\orcidlink{0000-0001-7888-6671}}
\email{schober@uni-bonn.de}
\affiliation{Argelander-Institut für Astronomie, Universität Bonn, Bonn 53121, Germany}

\date{\today}

\keywords{void magnetic fields, turbulent diffusivity, void galaxies, cosmological magnetic fields}

\begin{abstract}
Observations of TeV blazars imply a magnetic field that fills most of the intergalactic volume, including the voids of the large-scale structure.
The large-scale multipole fields of galaxies were recently proposed as a source of this field, but whether they reach far enough into the voids through (turbulent) diffusion is debated.
Here we quantify this reach by computing the volume that galactic fields fill in the observed population of voids, for the turbulent diffusivities proposed in the literature.
Our analytical field profile accounts for the 
profiles
of galactic multipole fields
found in simulations and stays above the bound out to several times the width of the diffusion front.
We find that galaxies can magnetize the voids to the level of the blazar bound within the published range of diffusivities.
The magnetized volume rises from insufficient to nearly complete over one decade in the diffusivity, which thus acts almost as a threshold.
\end{abstract}

\maketitle

%%%%%%%%%%%%%%%%%%%%%%%%%%%%%%%%%%%%%%%%%%%%%%%%%%%%%%%%%%%%%%
\section{Introduction}
\label{sec:intro}

Magnetic fields are observed in all gravitationally bound structures of the Universe, with strengths of order ten microgauss in galaxies \cite{ref:beck15} and a few microgauss in galaxy clusters \cite{ref:bonafede10,ref:donnert18}.
The non-observation of secondary $\gamma$-ray cascades from TeV blazars implies that a magnetic field of at least $10^{-17}$ to $10^{-16}\,$G is also present in the intergalactic medium \cite{ref:neronov10b,ref:acciari23}, where it fills at least 60\% of the volume \cite{ref:dolag11,ref:tjemsland24}.
Since the voids of the large-scale structure occupy most of the volume of the Universe \cite{ref:cautun14,ref:libeskind18}, this field must be present in them, far from the galaxies and clusters in which dynamos are known to operate \cite{ref:brandenburg23,ref:donnert18}.
The origin of the void field
is an open problem \cite{ref:durrer13,ref:subramanian16,ref:vachaspati21,ref:batista21}: either the field is primordial, generated in the early Universe, or it is astrophysical, generated in galaxies and spread from them into the voids.

Magnetic fields generated in the early Universe, during inflation or at cosmological phase transitions, permeate all of space and are therefore a natural explanation for the void field \cite{ref:durrer13,ref:vachaspati21}.
No mechanism is known, however, that generates fields of the required strength and coherence length within the Standard Model of particle physics and without fine-tuning.
Inflationary scenarios must break the conformal invariance of electromagnetism, which requires new couplings \cite{ref:ratra92,ref:anber10}, and are constrained by strong coupling and backreaction \cite{ref:demozzi09,ref:voneckardstein23,ref:voneckardstein25}; scenarios based on phase transitions require a first-order transition \cite{ref:sigl97,ref:ellis19}, whereas the electroweak and QCD transitions of the Standard Model are crossovers \cite{ref:donofrio16,ref:aoki06}.

The alternative is an astrophysical origin, in which fields amplified by dynamos in galaxies are transported into the intergalactic medium.
Galactic dynamos amplify weak seed fields to microgauss strengths within standard magnetohydrodynamics \cite{ref:schober13,
ref:brandenburg23,ref:subramanian16,ref:martinalvarez18}, and ordered fields are observed in galaxies at redshifts above two \cite{ref:geach23,ref:chen24,ref:deroo25}.
To account for the void field, however, these galactic fields must reach megaparsecs into the interiors of the voids.
The channel studied most is the transport of magnetized gas by galactic winds and by the outflows of active galactic nuclei \cite{ref:roychowdhury22,ref:samui18,ref:furlanetto01,ref:bertone06,ref:aramburogarcia21}.
Semi-analytic wind models yield field strengths of $10^{-12}$ to $10^{-8}\,$G in most of the regions reached by the winds \cite{ref:bertone06}, and in cosmological simulations the gas affected by outflows carries fields several orders of magnitude above those of unaffected gas at the same density \cite{ref:aramburogarcia21}.
The volume magnetized in this way depends strongly on how the outflows are modeled \cite{ref:bertone06}.
In several cosmological simulations, the magnetized regions remain close to the halos \cite{ref:katz19} and fill too small a fraction of the volume to account for the blazar observations \cite{ref:bondarenko22,ref:tjemsland24}, whereas outflows driven by cosmic rays extend to megaparsec scales \cite{ref:hopkins21,ref:samui18} and, in semi-analytic models, magnetize a volume sufficient to account for these observations \cite{ref:samui18}.
The feedback prescriptions that separate these outcomes are not yet constrained at the level required to decide between them \cite{ref:ayromlou23,ref:hadzhiyska25}.

In \Cite{my:dipole}, we proposed a different channel.
The multipoles of the large-scale galactic field extend beyond the galaxy, dominated at large distances by the lowest multipole, which we considered to be a dipole, falling off as the inverse cube of the distance. There, we show that the dipole field of a galaxy remains above the blazar bound out to distances of tens of megaparsecs, comparable to the sizes of voids.
This channel involves no transport of magnetized gas and therefore does not rely on the feedback prescriptions that govern the outflow channel, although galactic feedback can indeed contribute by driving additional turbulence in the void gas (see below).
Subsequent studies analyzed how such a magnetic field propagates into voids \cite{ref:seller26,ref:ghosh26,ref:brandenburg26,ref:yamashita26}.
Since this gas is a plasma, the magnetic field of a galaxy spreads into it by diffusion.
Within the diffusion front, the field is that of the multipole, and the distance covered by the front from the galaxy is set by the magnetic diffusivity of void gas and the time available \cite{ref:seller26,ref:ghosh26,ref:brandenburg26}.
For Spitzer resistivity, the front does not leave the galaxy \cite{ref:seller26}, but turbulent diffusion leads to a substantially larger spatial extent.
With order-of-magnitude estimates of the turbulent diffusivity of void gas, the diffusion front extends a few hundred kiloparsecs, which was judged insufficient to fill the voids \cite{ref:ghosh26,ref:brandenburg26}, whereas the diffusivity obtained from the turbulent velocities and scales of intergalactic gas in the TNG50 simulation gives a diffusion length of several megaparsecs, a significant fraction of the size of voids \cite{ref:yamashita26}.

Here,
we calculate the region around a galaxy that can be considered sufficiently magnetized and determine the total void volume that galactic fields
fill, as a function of diffusivity.
The field of a galaxy stays above the blazar bound out to a radius $\rth$, which we take as the reach of galactic fields.
Since the field of the galactic disk is far stronger than the bound, $\rth$ is several times the distance covered by the diffusion front.
Cosmic expansion extends it further by stretching the comoving distance over which the field diffuses.
To capture both effects, we derive the Gaussian profile of the field beyond the front from the analytical solution of \Cite{ref:seller26} and also show that it accounts for the profiles obtained in the simulations of \Cites{ref:ghosh26,ref:brandenburg26}.

The volume that the magnetized regions fill depends on where the galaxies lie: 
they are positioned
not only on the walls that enclose the voids but also, more sparsely, in their interiors \cite{ref:roychowdhury22}.
We therefore combine the Gaussian profile of the galactic field with the distribution of void sizes and the density of galaxies inside the voids measured in the Sloan Digital Sky Survey \cite{ref:pan12,ref:moorman15}.
Comparing the magnetized volume with the fraction implied by the blazar observations, we obtain the required diffusivity and find that the turbulent diffusivities adopted in \Cites{ref:ghosh26,ref:brandenburg26} can meet it and that obtained in \Cite{ref:yamashita26} exceeds it.
More importantly, the magnetized volume rises from insufficient to nearly complete over one decade in the diffusivity, which thus acts almost as a threshold.

The turbulent diffusivity of void gas has not been measured, but observations constrain a related quantity, \ie the diffusivity of cosmic rays in the halos of galaxies \cite{ref:butsky23,ref:yamasaki24,tex:crocker26}.
Since the turbulence carries the cosmic rays as it carries the field, their diffusivity tracks the turbulent diffusivity up to factors of order unity \cite{ref:krumholz20,ref:sampson23,tex:crocker26}.
The cosmic-ray diffusivity rises with distance from the galaxy, and the lower limits obtained in the outer halo already reach the threshold \cite{ref:butsky23}.
Farther out, an independent estimate of the diffusivity of intergalactic gas \cite{ref:yamashita26}, obtained from the TNG50 simulation, agrees with these limits.
The sensitivity of the magnetized volume to the diffusivity makes it imperative to determine the effective diffusivity of void gas including its factors of order unity.
Like Galactic emission for the cosmic microwave background, galaxy-seeded void fields would act as a foreground that has to be modeled before the void field can be used to constrain early-Universe physics.

This article is organized as follows.
\Cref{sec:prelim} describes the setup and defines the effective diffusivity.
\Cref{sec:spread} derives the profile of a galactic field spreading in an expanding universe and compares it with simulations.
\Cref{sec:filling} converts this profile to compute the magnetized region and the volume filled by it in a void.
\Cref{sec:fiducial} specifies the fiducial model and its inputs, and \cref{sec:requirement} gives the required diffusivity, its range, and its comparison with estimates and observations.
We discuss the assumptions and observational tests in \cref{sec:discussion} and conclude in \cref{sec:conclusions}.

%%%%%%%%%%%%%%%%%%%%%%%%%%%%%%%%%%%%%%%%%%%%%%%%%%%%%%%%%%%%%%
\section{Preliminaries}
\label{sec:prelim}

%-------------------------------------------------------------
\subsection{Setup}
\label{sec:setup}

\begin{figure}[t]
\includegraphics[width=0.95\columnwidth]{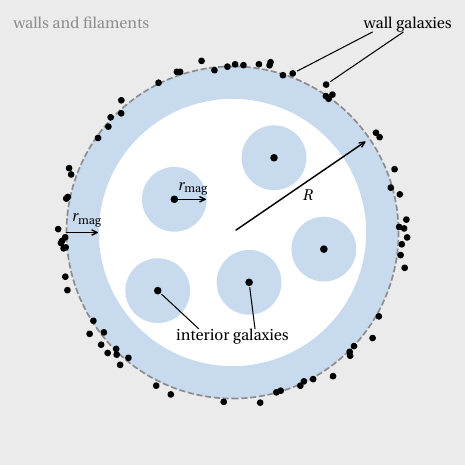}
\caption{
Schematic of the setup. A void of radius $R$ is enclosed by walls and filaments (gray), which we count as magnetized. Galaxies on the walls magnetize a shell of thickness $\rth$ along the edge of the void, and galaxies in the interior magnetize spheres of radius $\rth$ around themselves (blue).
}
\label{fig:schematic}
\end{figure}

We model a void as a sphere of effective radius $R$ enclosed by walls and filaments (\cref{fig:schematic}).
These host a large number of galaxies, close enough together for their fields to overlap, and we therefore count them as magnetized.
Two populations of galaxies contribute to the field inside the void: those on its walls and, at a much lower density, those in its interior.
We assume that each galaxy
hosts a large-scale field of microgauss strength, which far from the galaxy is dominated by its lowest multipole \cite{my:dipole,ref:seller26,ref:ghosh26,ref:brandenburg26}; we take the dipole as the fiducial geometry and compare it with quadrupolar fields (\cref{sec:sources}).
This field spreads into the surrounding gas by diffusion, at a rate set by the diffusivity of the gas (\cref{sec:spread}).
The field today stays above the blazar bound $\Bmin$ out to a radius $\rth$.
Each galaxy thus magnetizes a sphere of radius $\rth$ around itself (\cref{sec:reach}).
On the walls, where neighboring galaxies are closer than $\rth$ for the diffusivities of interest, these spheres merge into a continuous shell that extends $\rth$ into the void, whereas the spheres of the sparse interior galaxies remain mostly separate (\cref{fig:schematic}).
The magnetized volume is the union of the shell and the spheres, together with the walls and filaments (\cref{sec:coverage}).
We determine the diffusivity for which this volume reaches the fraction of space required by the blazar observations (\cref{sec:requirement}).

%-------------------------------------------------------------
\subsection{Effective induction equation}
\label{sec:eta}

We first consider a static background; the Hubble expansion is included in \cref{sec:comoving}.
In a gas of velocity $\vec v$, the magnetic field $\bb$ obeys the induction equation
\begin{gather}
    \pd_t \bb
    = \del \times \qty(\vec v \times \bb - \eta\sub{S} \mk \del \times \bb) \,,
    \label{eq:mhd}
\end{gather}
where $\eta\sub{S}$ is the microphysical diffusivity, which can be given by the Spitzer resistivity.
In a turbulent gas, we separate $\bb$ and $\vec v$ into averages over scales large compared to those of the turbulent fluctuations, $\bb = \ev{\bb} + \bb'$ and $\vec v = \ev{\vec v} + \vec v'$.
Averaging \cref{eq:mhd} according to Reynolds' averaging rules gives
\begin{gather}
    \pd_t \ev{\bb}
    = \del \times \qty(\ev{\vec v} \times \ev{\bb} + \vec{\mc E} - \eta\sub{S} \mk \del \times \ev{\bb}) \,,
    \label{eq:meanfield}
\end{gather}
in which the fluctuations enter through the mean electromotive force $\vec{\mc E} \equiv \ev{\vec v' \times \bb'}$.
For isotropic turbulence and to lowest order in the gradients of the mean field, $\vec{\mc E}$ can be modeled as
\cite{book:krause80,ref:brandenburg05},
\begin{gather}
    \vec{\mc E} = \alpha \mk \ev{\bb} - \eta\sub{t} \mk \del \times \ev{\bb} \,,
    \qquad
    \eta\sub{t} \simeq \frac{v \mk \ell}{3} \,,
    \label{eq:emf}
\end{gather}
where $v$ and $\ell$ are the velocity and the size of the largest turbulent eddies.
With this closure, \cref{eq:meanfield} becomes
\begin{gather}
    \pd_t \ev{\bb}
    = \del \times \qty[\ev{\vec v} \times \ev{\bb} + \alpha \mk \ev{\bb} - \qty(\eta\sub{S} + \eta\sub{t}) \mk \del \times \ev{\bb}] \,.
    \label{eq:meanind}
\end{gather}
The turbulent diffusivity thus adds to the microphysical one, and we define the effective diffusivity $\eta \equiv \eta\sub{S} + \eta\sub{t}$.
Since the void field lies many orders of magnitude below equipartition with the gas motions, the turbulence is strongly super-Alfv\'enic, and turbulent reconnection allows magnetic flux to be transported at rates set by the flow and independent of the microphysical resistivity \cite{ref:lazarian99,ref:eyink11}.
For the published estimates of void gas (\cref{sec:estimates}), $\eta\sub{t}$ exceeds $\eta\sub{S}$ by many orders of magnitude, so that $\eta \simeq \eta\sub{t}$.

Outside the galaxies, we neglect the mean flow and also set $\alpha = 0$, since the helical turbulence that the $\alpha$-effect requires arises from rotation and stratification inside the galaxies, where it drives the dynamo.
The galaxy then enters only as a prescribed source: we represent the currents maintained by its dynamo by the magnetization current $c \mk \del \times \vec M$, where the magnetization $\vec M$ vanishes outside the galaxy (\cref{sec:sources}).
From here on, $\bb$ denotes the mean magnetic field.
For a diffusivity that is uniform in space, the field then obeys \cite{ref:jedamzik11,ref:seller26}
\begin{gather}
    \pd_t \bb
    = \eta \qty[ \nabla^2 \bb + 4\pi \mk \del \times \qty(\del \times \vec M) ]
    = \eta \mk \nabla^2 \qty(\bb - \bb\sub{vac}) \,,
    \label{eq:static}
\end{gather}
where $\bb\sub{vac}$, with $\del \times \bb\sub{vac} = 4\pi \mk \del \times \vec M$, is the field that these currents would produce in vacuum.
The field thus relaxes toward the vacuum field of the galaxy at a rate set by $\eta$.
Inside the galaxy, the dynamo determines $\vec M$ along with the field; prescribing it instead fixes the field outside the galaxy.

%-------------------------------------------------------------
\subsection{Conventions}
\label{sec:conventions}

Throughout, we use Gaussian units and a spatially flat $\Lambda$CDM cosmology with $H_0 = \SI{67.7}{km.s^{-1}.Mpc^{-1}}$ and $\Omega_\mathrm{m} = 0.31$ \cite{ref:aghanim20}, for which the age of the Universe is $t\sub{H} = \SI{13.8}{Gyr}$.
We quote diffusivities in units of \mpckms, where $\SI{1}{\mpckms} \approx \SI{3.1e29}{cm^2.s^{-1}}$.
The inputs, their fiducial values, and the ranges over which we vary them are collected in \cref{tab:inputs} and discussed in \cref{sec:fiducial}.

%%%%%%%%%%%%%%%%%%%%%%%%%%%%%%%%%%%%%%%%%%%%%%%%%%%%%%%%%%%%%%
\section{Field spreading in an expanding universe}
\label{sec:spread}

%-------------------------------------------------------------
\subsection{Comoving formulation}
\label{sec:comoving}

We work in comoving coordinates $\vec x$ and conformal time $\tau$, with the scale factor $a = (1+z)^{-1}$, where $z$ is the redshift.
Further, $\nabla$ denotes from now on the gradient with respect to $\vec x$ and relates to the physical gradient through $\nabla^\mathrm{phys} = \nabla/a$.
Magnetic flux through a comoving surface is conserved in the absence of diffusion, so the natural variables are the comoving field $\bb \equiv a^2 \bb^{\rm phys}$ and the comoving magnetization $\vec M \equiv a^2 \vec M^{\rm phys}$, where the superscript denotes physical quantities.
In these variables, \cref{eq:static} keeps its form, with $t$ replaced by $\tau$ and $\eta$ by $\eta/a$ \cite{ref:jedamzik11,ref:seller26,my:chiralmhd_theory},
\begin{gather}
    \pd_\tau \bb = \frac{\eta}{a} \qty[ \nabla^2 \bb + 4\pi \mk \del \times \qty(\del \times \vec M) ] \,.
    \label{eq:induction}
\end{gather}

The scale factor enters only through the coefficient $\eta/a$, which can be absorbed into a diffusion `time' $s$,
\begin{gather}
    \dd s \equiv (\eta/a) \dd \tau = (\eta/a^2) \dd t  = - \eta(z)\dfrac{1+z}{H(z)}~\mathrm{d}z \,,
    \label{eq:ds}
\end{gather}
in terms of which \cref{eq:induction} becomes the diffusion equation in a static background with unit diffusivity [cf. \cref{eq:static}],
\begin{gather}
    \pd_s \bb
    = \nabla^2 \bb + 4\pi \mk \del \times \qty(\del \times \vec M) \,.
    \label{eq:diffusion}
\end{gather}
Here $s$ plays the role of time and carries the dimension of a squared length. In Fourier space, \cref{eq:diffusion} reads
\begin{gather}
    \pd_s \bb_{\vec k}
    = - k^2 \mk \bb_{\vec k} + 4\pi k^2 \vec M_{\vec k}^\perp \,,
    \label{eq:diffk}
\end{gather}
where $\perp$ denotes the component perpendicular to $\vec k$.

We take the source to be switched on at a redshift $z_i$ of order unity and hold $\vec M$ at its present value from then on.
For fixed $\vec M$, the comoving vacuum field of the galaxy is constant in time and falls off at large distances as $r^{-n}$, with $n = 3$ for a dipole.
In terms of the physical distance $r^{\rm phys} = a r$, the physical field is then $B^{\rm phys} = B/a^2 \propto 1/(a^2 r^n) = a^{n-2}/(r^{\rm phys})^n$.
A bound galaxy does not expand, so its edge stays at a fixed physical distance, and the field there scales as $a^{n-2}$.
For $n \geq 3$, a fixed $\vec M$ therefore gives a galactic field that was weaker in the past, whereas a field constant in physical units would require a source larger in the past by a factor $a^{2-n}$.
Holding $\vec M$ fixed is thus conservative for a galaxy whose field is approximately constant in physical units.

The diffusion time accumulated since redshift $z_i$ follows from \cref{eq:ds} as
\begin{gather}
    \ds (z_i) \equiv \int_0^{z_i} \eta(z) \mk \frac{1+z}{H(z)} \mk \dd z \,,
    \label{eq:clock}
\end{gather}
which coincides with the square of the inverse critical wavenumber of \Cite{ref:jedamzik11}.
The field today is then \cite{ref:jedamzik11,ref:seller26}
\begin{gather}
    \bb_{\vec k} = 4\pi \qty[1 - e^{-k^2 \ds (z_i)}] \vec M^\perp_{\vec k} \,.
    \label{eq:fourier}
\end{gather}
Since $\eta(z)$ and $H(z)$ enter only through $\ds$, \cref{eq:fourier} holds for any expansion history and any time dependence of the diffusivity.\footnote{The convergence condition of Ref.~\cite[Eq.~(9)]{ref:seller26} is required only for an eternal source, $z_i \to \infty$, for which the integral in \cref{eq:clock} diverges for a constant diffusivity in a matter-dominated Universe.} We separate the two by writing
\begin{gather}
    \ds (z_i) = q\sub{H} (z_i) \mk \bar\eta \mk t\sub{H} \,,
    \label{eq:dsfactor}
\end{gather}
where
\begin{gather}
    q\sub{H} (z_i) \equiv \frac{1}{t\sub{H}} \int_0^{z_i} \frac{1+z}{H(z)} \mk \dd z
    \label{eq:qh}
\end{gather}
depends only on the expansion history, and \cref{eq:dsfactor} defines $\bar\eta$, the average of $\eta(z)$ over $0 < z < z_i$ weighted by $(1+z)/H(z)$.
The factor $q\sub{H}$ is thus the ratio of the accumulated diffusion time to the static estimate $\bar\eta \mk t\sub{H}$ over the age of the Universe.
For the $\Lambda$CDM model adopted here (\cref{sec:conventions}), $q\sub{H} = 1.74$, $2.28$, and $2.88$ for $z_i = 1.5$, $2$, and $2.6$.
It exceeds unity although the source acts for less than the age of the Universe, because a physical displacement made when $a$ was smaller corresponds to a larger comoving distance.

%-------------------------------------------------------------
\subsection{Galactic source}
\label{sec:sources}

The magnetization $\vec M$ represents the maintained source that is obtained from the galactic dynamos, which are confined to the galaxy, $r \lesssim r\sub{G}$.
Since the galactic radius, of order ten kiloparsecs, is small compared to both the distance over which the field diffuses and the distances at which the void field matters, of order a megaparsec, the field there depends only on the lowest multipole moments of $\vec M$, and we replace the galaxy by a point source.
For a dipole of constant comoving moment $\vec m$, whose vacuum field falls off as $r^{-3}$, we obtain
\begin{gather}
    \vec M = \vec m \mk \delta^3(\vec x) \,,
    \quad
    \vec M_{\vec k} = \vec m \,.
    \label{eq:Mdip}
\end{gather}
With the heat kernel $G_\sigma(r) = (4\pi\sigma)^{-3/2} \mk e^{-r^2/4\sigma}$, where $r = \abs{\vec x}$ is the comoving distance from the galaxy, the solution of \cref{eq:diffusion} for this source is
\begin{gather}
    \bb(\vec r) = 4\pi \int_0^{\ds} \del \times \qty[\del \times \qty(\vec m \mk G_\sigma)] \dd \sigma \,,
    \label{eq:realspace}
\end{gather}
in which the contribution emitted a diffusion time $\sigma$ ago has spread into the Gaussian $G_\sigma$.
Evaluating the integral, or equivalently transforming \cref{eq:fourier} to real space, gives \cite{ref:seller26}
\begin{gather}
    \bb(\vec r)
    = \frac{3\mk \vu r \mk (\vec m \cdot \vu r) \mk \mc F(u;6) - \vec m \mk \mc F(u;2)}{r^3} \,,
    \quad u \equiv \frac{r}{\sqrt{\ds}} \,,
    \label{eq:exact}
\end{gather}
with
\begin{gather}
    \mc F(u;n) = 1 - \operatorname{erf}\qty(\frac{u}{2}) + \frac{u}{\sqrt{\pi}} \mk e^{-u^2/4} \qty(1 + \frac{u^2}{n}) \,.
    \label{eq:Fs}
\end{gather}
For $u \ll 1$, $\mc F \to 1$ and \cref{eq:exact} reduces to the vacuum dipole.
For $u \gg 1$,
\begin{gather}
    \mc F(u;n) \simeq \frac{u^3}{n\sqrt{\pi}} \mk e^{-u^2/4} \,,
    \label{eq:farF}
\end{gather}
and the two terms combine into
\begin{gather}
    \bb(\vec r) \simeq - \frac{\vec m_\perp}{2\sqrt{\pi} \mk (\ds)^{3/2}} \mk e^{-r^2/(4\ds)} \,,
    \label{eq:farfield}
\end{gather}
where $\vec m_\perp \equiv \vec m - \vu r \mk (\vu r \cdot \vec m)$ is the component of the moment perpendicular to the radial direction.
The field is then purely transverse and Gaussian in $r$: the growth of $\mc F$ cancels the $r^{-3}$ falloff of the dipole.
We define the front width as the standard deviation of this Gaussian,
\begin{gather}
    \rs \equiv \sqrt{2 \ds} \,,
    \qquad \rs^2 = 2 \mk q\sub{H} \mk \bar\eta \mk t\sub{H} \,,
    \label{eq:rstar}
\end{gather}
which is the root-mean-square displacement per dimension of Brownian motion over the diffusion time $\ds$.

\begin{figure}[t]
\includegraphics[width=\columnwidth]{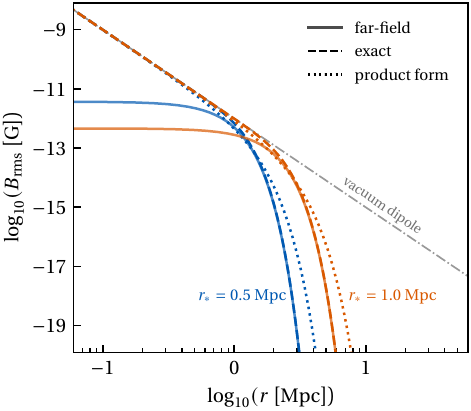}
\caption{RMS field of a sustained dipole of strength $B\sub{G} = 1\,\mu\mathrm{G}$ at $r\sub{G} = \SI{10}{kpc}$, for two front widths. Solid lines show the far-field model~\eqref{eq:fiducial}, dashed lines the exact form~\eqref{eq:brms} with the same $\rs$, and the dotted line the product form~\eqref{eq:prod} with $r\sub{prod} = \sqrt{5\mk\ds}$. The faint gray dotted line is the RMS field of the vacuum dipole.
}
\label{fig:profile}
\end{figure}

%-------------------------------------------------------------
\paragraph*{RMS profile.}
We characterize the field at distance $r$ by its root-mean-square (RMS) over directions,
\begin{gather}
    B\sub{rms} (r) \equiv \qty(\frac{1}{4\pi} \int \abs{\bb (\vec r)}^2 \dd \Omega)^{1/2} \,,
    \label{eq:rms}
\end{gather}
which for the exact solution~\eqref{eq:exact} equals
\begin{subequations}
\label{eq:brms}
\begin{gather}
    B\sub{rms}(r) = \frac{m}{r^3} \mk g(u) \,,
    \\
    g(u) \equiv \sqrt{\frac{1}{3} \qty[ 3 \mc F(u;6) - \mc F(u;2) ]^2 + \frac{2}{3} \mk \mc F(u;2)^2 } \,,
\end{gather}
\end{subequations}
where $m \equiv \abs{\vec m}$.
The two terms are the radial and transverse components, weighted by their angular averages $1/3$ and $2/3$.
We normalize the solution through the RMS galactic field $B\sub{G}$ at the galactic radius $r\sub{G}$.
Since $r\sub{G} \ll \sqrt{\ds}$, the field there is that of the vacuum dipole, for which $g \to \sqrt{2}$, so that $B\sub{G} = \sqrt{2} \mk m/r\sub{G}^3$.
Beyond the front, \cref{eq:farfield} gives
\begin{gather}
    B\sub{rms}(r) \simeq \frac{m \mk e^{-r^2/2\rs^2}}{\sqrt{6\pi} \mk (\ds)^{3/2}}
    = \sqrt{\frac{2}{3\pi}} \mk B\sub{G} \qty(\frac{r\sub{G}}{\rs})^3 e^{-r^2/2\rs^2} \,.
    \label{eq:fiducial}
\end{gather}
Since the magnetized region defined in \cref{sec:reach} lies several front widths out, where the two profiles merge (\cref{fig:profile}), we use this far-field model hereafter; it reproduces the reach of the exact solution to better than a percent.

%-------------------------------------------------------------
\paragraph*{Comparison with simulations.}
The moment radius of the exact profile~\eqref{eq:brms}, $r\sub{M} \equiv \int B\sub{rms} \mk r^4 \dd r / \int B\sub{rms} \mk r^3 \dd r$, is given by $r\sub{M}^2 = 3.185 \mk \ds$, consistent with the simulations of \Cite{ref:ghosh26}, in which it exceeded $1.7 \mk \ds$ and was still rising at the end of the runs.
Simulated profiles are commonly described by the product form
\begin{gather}
    B^{\rm prod} (r) = \frac{\sqrt{2} \mk m}{r^3} \exp(-r^2/2 r\sub{prod}^2) \,,
    \label{eq:prod}
\end{gather}
the RMS vacuum dipole with a Gaussian cutoff, whose moment radius is $\sqrt{2/\pi} \mk r\sub{prod}$.
Matching it to that of the exact profile gives $r\sub{prod}^2 = (\pi/2) \mk r\sub{M}^2 = 5 \mk \ds$, which accounts for the value $r\sub{prod}^2 \approx 4.9 \mk \ds$ fitted to the simulated profiles in \Cite{ref:brandenburg26}.
This width exceeds the Brownian value $\rs^2 = 2 \mk \ds$ because the product form retains the $r^{-3}$ factor beyond the front, where the exact field~\eqref{eq:farfield} has lost it.
Its Gaussian is wider than that of \cref{eq:fiducial} by a factor $(5/2)^{1/2} \approx 1.6$, so that the product form reaches farther despite the $r^{-3}$ factor. The far-field model~\eqref{eq:fiducial} used here is the more conservative of the two (\cref{fig:profile}).

%-------------------------------------------------------------

%-------------------------------------------------------------
\paragraph*{Quadrupolar fields.}
Galactic dynamos can generate fields that are either antisymmetric or symmetric about the galactic plane, known as dipolar and quadrupolar parity \cite{ref:brandenburg23,ref:qazi25}.
We therefore also consider the two components of a quadrupolar field: the poloidal quadrupole, whose vacuum field falls off as $r^{-4}$, and the toroidal component found in simulations of quadrupolar dynamos, which falls off as $r^{-2}$ \cite{ref:brandenburg26}.
We model their RMS far fields by \cref{eq:fiducial} with the power of $r\sub{G}/\rs$ set by the respective falloff,
\begin{gather}
    B\sub{rms}^{\rm pol}(r) \simeq \sqrt{\frac{2}{3\pi}} \mk B\sub{G} \qty(\frac{r\sub{G}}{\rs})^n e^{-r^2/2\rs^2} \,,
    \label{eq:quadfields}
\end{gather}
where $n=4$ for the poloidal quadrupolar field and $n=2$ for the toroidal quadrupolar field. All three geometries are assigned the same strength $B\sub{G}$ at $r\sub{G}$, since the saturation strength of a galactic dynamo is set by equipartition with the turbulent kinetic energy of the interstellar gas, independently of the parity of the field \cite{ref:brandenburg23,ref:beck15}.

%-------------------------------------------------------------
\paragraph*{Start redshift and source history.}
The start redshift $z_i$ marks the epoch by which galaxies acquired large-scale fields comparable to their present ones.
A galactic field that reaches its present strength only later is described by a lower $z_i$.
A time-dependent moment enters \cref{eq:realspace} under the integral, evaluated at the time of emission.
The comoving moment could have been larger in the past, as for a bound galaxy of constant physical moment $\vec m^{\rm phys}$.
For simplicity, we hold it at its present value from $z_i$ on. This is a conservative choice (\cref{sec:comoving}).

%%%%%%%%%%%%%%%%%%%%%%%%%%%%%%%%%%%%%%%%%%%%%%%%%%%%%%%%%%%%%%
\section{Volume filling}
\label{sec:filling}

%-------------------------------------------------------------
\subsection{Magnetized region}
\label{sec:reach}

We define the radius $\rth$ of the magnetized region by $B\sub{rms}(\rth) = \Bmin$, where the threshold $\Bmin$ is set by the blazar bound \cite{ref:acciari23}; for the profiles \eqref{eq:fiducial} and \eqref{eq:quadfields}, this gives
\begin{gather}
    \rth = \rs \sqrt{2 \Lambda} \,,
    \qquad
    \Lambda \equiv \ln\qty[\sqrt{\frac{2}{3\pi}} \frac{B\sub{G}}{\Bmin} \qty(\frac{r\sub{G}}{\rs})^n] \,,
    \label{eq:rth}
\end{gather}
where $\rs = \sqrt{2 \mk q\sub{H} \mk \bar\eta \mk t\sub{H}}$, and $n=2, 3, \text{ and } 4$ for a toroidal quadrupole field, a dipole field, and a poloidal quadrupole field [see \cref{eq:fiducial,eq:quadfields}], respectively.
\Cref{eq:rth} relies on the far-field form, which requires $\Lambda \gg 1$; for the parameters considered here, $\rth$ lies several front widths out. We assume that $\Lambda$ is sufficiently high that we can assume monotonic increase of $\rth$ with respect to $\rs$. For the fiducial model considered below, $\Lambda \simeq 10.5$ when the error in the calculated $\rth$ is less than $0.25\%$. We observe empirically that the far-field formula holds well until $\Lambda \sim 4$ with the error in the calculated magnetic field being less than~$10\%$ (\cref{fig:profile}).

The reach is linear in the front width and only logarithmic in the other parameters.
 The field amplitudes enter only through $\Lambda$, so a factor of ten in $B\sub{G}$, in $\Bmin$, or in the normalization of the profile changes the reach by only about ten percent for the fiducial model described below in \cref{tab:inputs,tab:fiducial}.

At a fixed distance, by contrast, the field depends steeply on the front width.
At the surface of the magnetized region, the Gaussian factor in \cref{eq:fiducial} equals $e^{-\Lambda}$, and since its exponent scales as $\rs^{-2}$, the response to a finite change of $\rs$ is strongly asymmetric: tripling $\rs$ raises the field there by a factor $e^{8\Lambda/9}/3^n$, while reducing it to a third lowers the field by a factor $e^{8\Lambda}/3^n$.

For the fiducial model, with $\Lambda \simeq 10.5$ and $n = 3$ (\cref{tab:fiducial}), these amount to three and thirty-five orders of magnitude, respectively.
Factors of order unity in the front width are therefore crucial for determining how much of a void is magnetized: among them are $q\sub{H}$ and the measure of the diffusion length itself, which in the literature ranges from $\sqrt{\ds}$ \cite{ref:jedamzik11,ref:seller26} to $2\pi\sqrt{\ds}$ \cite{ref:yamashita26}, against the front width $\rs = \sqrt{2\ds}$.

\begin{figure}[t]
\includegraphics[width=\columnwidth]{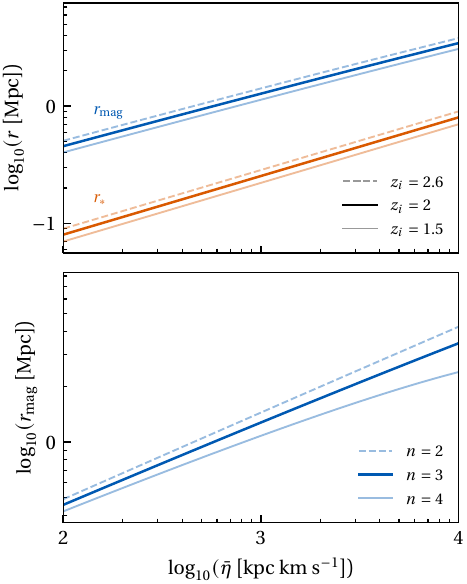}
\caption{Front width $\rs$~\eqref{eq:rstar} and the radius of the magnetized region $\rth$~\eqref{eq:rth} as functions of $\bar\eta$, with the other inputs at their fiducial values. Top: $\rs$ (orange) and $\rth$ (blue) of the dipole for three start redshifts. Bottom: $\rth$ of the dipole ($n = 3$), the poloidal quadrupole ($n = 4$), and the toroidal quadrupole ($n = 2$) for $z_i = 2$. Dark lines show the fiducial model and faint lines the ends of the ranges in \cref{tab:inputs}.}
\label{fig:reach}
\end{figure}

The top panel of \cref{fig:reach} shows $\rs$ and $\rth$ over the relevant range of diffusivities and start redshifts, and the bottom panel compares the three geometries.
The reach of the poloidal quadrupole flattens toward high $\bar\eta$: inside the front, the field follows the $r^{-4}$ falloff of the vacuum quadrupole, which reaches $\Bmin$ at a fixed distance, and once the front approaches that distance, a further increase of $\bar\eta$ leaves the reach unchanged.

%-------------------------------------------------------------
\subsection{Coverage of a void}
\label{sec:coverage}

Using the results from \cref{sec:reach} for the magnetized region, we can now evaluate how much volume of a given void can be filled for a given galaxy number density and radius of the magnetized region $\rth$.
Consider a void of radius $R$.
Galaxies on its boundary are spaced by about a megaparsec \cite{ref:seller26}, less than $\rth$ for the diffusivities of interest, so together they magnetize a shell of thickness $\rth$ along the walls (see \cref{fig:schematic} for a schematic). The shell from the wall galaxies fills the fraction
\begin{gather}
    f\sub{shell} (R) = 1 - \qty(1 - \frac{\rth}{R})^3 \,,
    \qquad R > \rth \,,
    \label{eq:shell}
\end{gather}
of the void volume.
Because most of the volume of a sphere lies close to its surface, a shell of modest thickness fills a large fraction of the void volume.
Galaxies in the void interior magnetize spheres of radius $\rth$ around themselves.
We take them to be placed independently at random with mean density $\nG$, that is, as a Poisson process.
A point is then uncovered if the sphere of volume $V\sub{mag} = 4\pi\rth^3/3$ centered on it contains no galaxy. Since a volume element $\delta V$ is occupied independently with probability $\nG\,\delta V$, all of them are empty with probability $(1 - \nG\,\delta V)^{V\sub{mag}/\delta V}$, which tends to $e^{-\nG V\sub{mag}}$ as $\delta V \to 0$.
The interior galaxies therefore cover the fraction
\begin{gather}
    f\sub{in} = 1 - \exp\qty(-\frac{4\pi}{3} \mk \nG \mk \rth^3) \,.
    \label{eq:poisson}
\end{gather}
For sparse galaxies, $\nG V\sub{mag} \ll 1$, this reduces to $\nG V\sub{mag}$, the total volume of the spheres per unit volume; the exponential accounts for their overlap.

Treating the two contributions as independent, the magnetized fraction of the void is
\begin{gather}
    f\sub{void} (R) = 1 - \qty[1 - f\sub{shell}(R)] \qty(1 - f\sub{in}) \,.
    \label{eq:fvoid}
\end{gather}
The volume outside the voids, occupied by walls, filaments, and halos, makes up a fraction $f\sub{s}$ of the total, which we take to be fully magnetized \cite{ref:vernstrom21}.
The total filling factor is then
\begin{gather}
    f\sub{tot} = f\sub{s} + (1 - f\sub{s}) \mk f\sub{void} \,.
    \label{eq:ftot}
\end{gather}
The two populations, the wall galaxies and the galaxies in the void interior, dominate in different regimes.
In small voids, the shell from the walls fills most of the volume, and $f\sub{void} \to f\sub{shell}$ as $R \to \rth^+$.
In large voids, the shell fills a decreasing fraction and $f\sub{void} \to f\sub{in}$ for $R \to \infty$.

%%%%%%%%%%%%%%%%%%%%%%%%%%%%%%%%%%%%%%%%%%%%%%%%%%%%%%%%%%%%%%
\section{Fiducial model}
\label{sec:fiducial}

The diffusivity required to obtain a sufficient filling factor for the magnetized part of the voids depends on a multitude of inputs that describe the voids, the galaxies on their walls and in their interiors, and the galactic fields.
In this section, we specify a fiducial model and the alternatives against which we test it, beginning with the void sizes.

%-------------------------------------------------------------
\begin{figure}[t]
\includegraphics[width=\columnwidth]{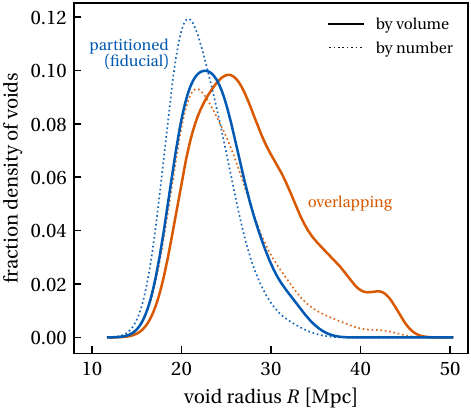}
\caption{
Distribution of void radii in the volume-limited SDSS DR7 sample, by volume, $f_V(R)$ (solid), and by number, $f_N(R)$ (dotted), for partitioned (fiducial) and overlapping volumes (\cref{sec:voidsize}). Both volume distributions are normalized to the total partitioned void volume, so that the excess of the overlapping curve shows the volume counted in more than one void; the number distributions are normalized to unity.
}
\label{fig:voidsize}
\end{figure}

%-------------------------------------------------------------
\begin{table*}
\begin{ruledtabular}
\begin{tabular}{lccl}
Input & Fiducial & Range & Source \\[2pt]
\hline \noalign{\vskip 4pt}
Void radius, $R$ [Mpc] & distribution in \cref{fig:voidsize} & 15 -- 35 & \cite{ref:elad97,ref:hoyle02,ref:douglass22} \\[2pt]
Void-galaxy density, $\nG$ [\si{Mpc^{-3}}] & \num{5e-3} & \num{2.5e-3} -- $10^{-2}$ & \cite{ref:hoyle05,ref:moorman15} \\[2pt]
Galactic field amplitude, $B\sub{G}$ [G] & \num{e-6} & \num{3e-7} -- \num{3e-6} & \cite{ref:beck15,ref:unger24} \\[2pt]
Radius of the galaxy, $r\sub{G}$ [kpc] & 7 & 3 -- 10 & \cite{ref:beck15,ref:unger24,ref:gaensler05} \\[2pt]
Lower bound, $\Bmin$ [G] & \num{3e-17} & $10^{-17}$--$10^{-16}$ & \cite{ref:acciari23} \\[2pt]
Filling requirement, $\freq$ & 0.65 & 0.6 -- 0.7 & \cite{ref:dolag11,ref:tjemsland24} \\[2pt]
Volume fraction outside voids $f\sub{s}$ & 0.35 & 0.25 -- 0.4 & \cite{ref:cautun14,ref:libeskind18,ref:douglass23} \\[2pt]
Start redshift, $z_i$ & 2 & 1.5 -- 2.6 & \cite{ref:mao17,ref:geach23,ref:chen24,ref:deroo25} \\[2pt]
Field geometry, $n$ & dipole ($n=3$) & poloidal ($n=4$) and toroidal ($n=2$) quadrupole & \cite{ref:brandenburg26}
\end{tabular}
\end{ruledtabular}
\caption{Inputs, their fiducial values, and the ranges over which they are varied and the references from which these inputs are drawn.}
\label{tab:inputs}
\end{table*}

%-------------------------------------------------------------
\subsection{Void size}
\label{sec:voidsize}

The magnetized fraction of a void depends on its size (\cref{sec:coverage}), so the total filling factor depends on how the void volume is distributed over sizes.
We therefore replace the single radius in \cref{eq:ftot} by a distribution of void sizes, taken from observations as a representative example of a realistic distribution.
Excursion-set models also predict such a distribution \cite{ref:sheth04}, but the prediction depends on how voids are defined and traced, and must be calibrated against simulations for each void finder \cite{ref:jennings13,ref:pisani15}.
The observed distribution we adopt uses a void definition that matches the division into walls and void interiors in \cref{eq:ftot}, and its voids are a subset of those in which the density of void galaxies was measured \cite{ref:pan12,ref:moorman15}.

We identify voids with VoidFinder \cite{ref:elad97,ref:hoyle02}, as implemented in the Void Analysis Software Toolkit (VAST) \cite{ref:douglass22}.
After separating isolated field galaxies by the distance to their third-nearest neighbor, VoidFinder grows the largest spheres empty of the remaining wall galaxies.
A void is a maximal sphere of radius at least $10\,h^{-1}\,$Mpc together with the smaller spheres that overlap it predominantly.
These voids end at the walls, as required by the division of the volume in \cref{eq:ftot} into walls and void interiors.
Watershed void finders instead extend voids to the crests of the walls, counting part of the wall volume as void \cite{ref:zaidouni25}.

We apply VoidFinder, with the default settings of VAST along with the values stated in \cref{sec:conventions}, to the volume-limited sample of the Sloan Digital Sky Survey Data Release 7 from \Cite{ref:pan12}, which contains the galaxies brighter than $M_r - 5\log h = -20.09$ out to $z = 0.1026$.
The minimum radius of the maximal spheres is nearly twice the mean galaxy separation in this sample, which makes the voids robust against shot noise and peculiar velocities \cite{ref:pisani15}.
Since \Cite{ref:pan12} built their void catalog from the same sample and \Cite{ref:moorman15} measured the void luminosity function in those voids, our void sizes and the galaxy density $\nG$ of \cref{sec:inputs} refer to a subset of the same void population.
The 883 voids we find occupy 67.5\% of the surveyed volume.

The volume of each void is obtained by counting the points of a uniform grid, with a spacing of $2\,h^{-1}\,$Mpc, that lie inside both the void and the surveyed region; a finer grid gives the same volumes.
Neighboring voids overlap, and about a third of the void volume is claimed by more than one void.
We assign each shared point to the void whose maximal sphere is centered nearest, so that the void volumes add up to the total void volume (``partitioned'' in \cref{fig:voidsize}).
Published catalogs often instead count the shared points in every void that contains them \cite{ref:pan12, ref:douglass23, tex:curtis26} (``overlapping'').
We characterize each void by its effective radius $R = (3V/4\pi)^{1/3}$, converted to Mpc with $h = 0.677$.
The effective radii lie between $10.3\,h^{-1}~{\rm Mpc} \, (\SI{15.2}{Mpc})$ and $29.5\,h^{-1}~{\rm Mpc} \,  (\SI{43.6}{Mpc})$ for overlapping volumes, the largest as in the catalog of \Cite{ref:pan12}, and between $10.1\,h^{-1}~{\rm Mpc} \, (\SI{14.9}{Mpc})$ and $23.5\,h^{-1}~{\rm Mpc} \,  (\SI{34.8}{Mpc})$ for partitioned volumes.

The void population is characterized by two distributions over the effective radius, each normalized to unity: $f_N(R)\,\dd R$ is the fraction of voids with radii between $R$ and $R + \dd R$, and $f_V(R)\,\dd R$ is the fraction of the void volume in such voids.
We obtain both as density estimates, summing Gaussians of width $1\,h^{-1}\,$Mpc centered on the void radii, with equal weights for $f_N$ and volume weights for $f_V$ (\cref{fig:voidsize}).
Since the filling factor is a fraction of volume, the relevant distribution is $f_V$, and \cref{eq:ftot} becomes
\begin{gather}
    f\sub{tot} = f\sub{s} + (1 - f\sub{s}) \int f\sub{void}(R) \mk f_V(R) \mk \dd R \,,
    \label{eq:ftotdist}
\end{gather}
where $f\sub{void}$ is given by \cref{eq:fvoid}, since all voids in the catalog are larger than the radius of the magnetized region for the diffusivities considered here.

\paragraph*{Alternatives.}
As alternatives, we replace the distribution by a single void size at either end of the catalog: all voids with $R = \SI{15}{Mpc}$, close to the minimum radius of VoidFinder voids, and all voids with $R = \SI{35}{Mpc}$, roughly the largest in the catalog.
Since $f\sub{void}$ decreases monotonically with $R$, any distribution of void sizes confined between these radii yields a required diffusivity between the two cases.

%-------------------------------------------------------------
\subsection{Other inputs}
\label{sec:inputs}

The remaining inputs are listed in \cref{tab:inputs}, together with their fiducial values and the ranges over which we vary them, one at a time with the others held fixed.
We motivate each of them below.

\paragraph*{Galaxy density in voids.}
The density of galaxies in the void interiors follows from the void-galaxy luminosity function, which \Cite{ref:moorman15} measured in the voids of \cref{sec:voidsize} and \Cite{ref:hoyle05} in an earlier SDSS sample.
Integrating it down to a limiting magnitude gives the densities in \cref{tab:ng}.
These densities likely underestimate the galaxy population relevant here: the Schechter fits from which they follow underestimate the faint end \cite{ref:moorman15}, low-surface-brightness dwarfs are undercounted at all magnitudes \cite{ref:moorman15}, and the densities are measured at $z \lesssim 0.1$, whereas for $z_i = 2$ about half of the diffusion time accumulates at $z > 1$, when voids were less evacuated \cite{ref:sheth04}.
We adopt $\nG = \SI{5e-3}{Mpc^{-3}}$, the measured density down to $M_r \approx -13$, and vary it by a factor of two in each direction; the lower end corresponds to galaxies brighter than $M_r \approx -15$.

%-------------------------------------------------------------
\begin{table}[t]
\begin{ruledtabular}
\begin{tabular}{lcc}
$M_r$ & $-15$ & $-13$ \\[2pt]
\hline \noalign{\vskip 4pt}
\citet{ref:hoyle05} & 4.5 & 7.8$^*$ \\[2pt]
\citet{ref:moorman15} & 2.6 & 4.8 \\
\end{tabular}
\end{ruledtabular}
\caption{Number density of void galaxies brighter than $M_r$, in units of $10^{-3}\,\si{Mpc^{-3}}$, from the void luminosity functions of \Cites{ref:hoyle05,ref:moorman15}, both measured at $z \lesssim 0.1$. The Schechter parameters of \Cite{ref:hoyle05} are converted with $h = 0.677$. $^*$Extrapolated beyond the faintest magnitude of their sample, $M_r = -14.5$.}
\label{tab:ng}
\end{table}
%-------------------------------------------------------------

\paragraph*{Galactic field.}
The profile of the galactic field is the far-field model~\eqref{eq:fiducial}.
We adopt $B\sub{G} = 1\,\mu$G at $r\sub{G} = \SI{7}{kpc}$.
Coherent fields of this strength are found across the galaxy population: large spirals host ordered fields of a few microgauss across their disks \cite{ref:beck15,ref:unger24}, and even the Large Magellanic Cloud, a gas-rich irregular galaxy, hosts a coherent spiral field of about $1\,\mu$G \cite{ref:gaensler05}.
We vary $B\sub{G}$ by a factor of three in each direction and $r\sub{G}$ between $3$ and $\SI{10}{kpc}$, spanning disks from half the size of the Large Magellanic Cloud to the typical size of large spirals.
Since the dipole moment enters the reach only through $\Lambda$ in \cref{eq:rth}, a factor of ten in the moment changes the reach by about ten percent.
The weaker ordered fields of faint dwarfs therefore shorten their reach only moderately, and the same galaxies are undercounted by the optical luminosity functions \cite{ref:moorman15}.

\paragraph*{Lower bound and filling requirement.}
For the lower bound we adopt $\Bmin = \SI{3e-17}{G}$.
It lies between the bound obtained under conservative assumptions on the activity of the sources, $\SI{1.8e-17}{G}$ for coherence lengths above $\SI{0.2}{Mpc}$ \cite{ref:acciari23}, and the higher bounds obtained under less conservative ones.
The fields computed here vary on scales of at least $\SI{0.2}{Mpc}$.
We vary $\Bmin$ between $10^{-17}$ and $10^{-16}\,$G.
The blazar observations require a volume filling factor between $0.60$ and $0.67$ \cite{ref:dolag11,ref:tjemsland24}.
We adopt $\freq = 0.65$ and vary it between $0.6$ and $0.7$.

\paragraph*{Volume of walls and filaments.}
The fraction of the volume occupied by walls, filaments, and halos, $f\sub{s}$, depends on where the void boundaries are drawn.
For the void definition of \cref{sec:voidsize}, the SDSS catalog gives $f\sub{s} = 0.325$, and the catalog of \Cite{ref:pan12}, in whose voids $\nG$ was measured \cite{ref:moorman15}, gives about 0.38; we adopt $f\sub{s} = 0.35$, between the two.
Other void catalogs of the SDSS give values between 0.32 and 0.39 \cite{ref:douglass23}.
Classifications of the cosmic web in simulations give values between 0.23 and 0.34 \cite{ref:cautun14,ref:libeskind18}; the lowest of these count part of the volume that lies in the walls as defined here.
We therefore vary $f\sub{s}$ between 0.25 and 0.40, the range plausible for voids that end at the walls.

\paragraph*{Start redshift.}
A coherent microgauss field is observed through Faraday rotation in a galaxy at $z = 0.44$ \cite{ref:mao17}, and polarized dust emission reveals fields ordered on kiloparsec scales in lensed galaxies at $z \approx 2.6$, where the field follows the spiral arms as in nearby spirals \cite{ref:geach23,ref:deroo25}, and at $z \approx 5.6$, where it appears shaped by a merger \cite{ref:chen24}.
Mean-field dynamo models likewise produce regular microgauss fields within about $2\,$Gyr of disk formation, by $z \approx 3$ \cite{ref:arshakian09}, so that ordered galactic fields arise early and through more than one channel.
We adopt $z_i = 2$ and vary it between $1.5$ and $2.6$, the redshift of the most distant field observed to follow spiral arms; with ordered fields observed out to $z \approx 5.6$, this range is conservative.

\paragraph*{Field geometry.}
The fiducial geometry is the dipole.
As alternatives, we use the poloidal and toroidal components of a quadrupolar dynamo field~\eqref{eq:quadfields}, between which the dipole lies in reach (\cref{fig:reach}). We vary this by changing $n \in \qty{2,3,4}$ in \cref{eq:quadfields}.

%-------------------------------------------------------------
\begin{table}[t]
\begin{ruledtabular}
\begin{tabular}{lcc}
Quantity & Eq. & Value \\[2pt]
\hline \noalign{\vskip 4pt}
% requirement and distances
$\etareq\,(f\sub{tot} = \freq)$        & \eqref{eq:ftotdist} & \SI{4.3}{\mpckms} \\[2pt]
$q\sub{H}$                             & \eqref{eq:qh}       & 2.28 \\[2pt]
$\rs$                                  & \eqref{eq:rstar}    & \SI{0.52}{Mpc} \\[2pt]
$\Lambda$                              & \eqref{eq:rth}      & 10.5 \\[2pt]
$\rth$                                 & \eqref{eq:rth}      & \SI{2.4}{Mpc} \\[2pt]
% coverage at etareq
$f\sub{shell}$                         & \eqref{eq:shell}    & 0.28 \\[2pt]
$f\sub{in}$                            & \eqref{eq:poisson}  & 0.25 \\[2pt]
$f\sub{void}$                          & \eqref{eq:fvoid}    & 0.46 \\[2pt]
% each population alone
$\etareq\,(\nG \to 0)$                 & \eqref{eq:shell}    & \SI{1.7e1}{\mpckms} \\[2pt]
$\etareq\,(R \to \infty)$              & \eqref{eq:poisson}  & \SI{7.7}{\mpckms} \\[2pt]
$\etareq\,(f\sub{s} = 0)$              & \eqref{eq:fvoid}    & \SI{7.8}{\mpckms} \\[2pt]
% steepness
$\bar\eta\,(f\sub{tot} = 0.5)$         & \eqref{eq:ftotdist} & \SI{1.4}{\mpckms} \\[2pt]
$\bar\eta\,(f\sub{tot} = 0.9)$         & \eqref{eq:ftotdist} & \SI{1.4e1}{\mpckms} \\
\end{tabular}
\end{ruledtabular}
\caption{Derived quantities for the fiducial model. Distances and coverages are evaluated at $\etareq$.}
\label{tab:fiducial}
\end{table}

%%%%%%%%%%%%%%%%%%%%%%%%%%%%%%%%%%%%%%%%%%%%%%%%%%%%%%%%%%%%%%
\section{Required diffusivity}
\label{sec:requirement}

We define the minimum required diffusivity $\etareq$ as the value of $\bar\eta$ at which the total filling factor~\eqref{eq:ftotdist} reaches $\freq$.

%-------------------------------------------------------------
\subsection{Fiducial model}
\label{sec:req_fiducial}

For the fiducial inputs, $\etareq \approx \num{4.3}\,\mpckms$ (\cref{tab:fiducial}).
The front is then about half a megaparsec wide, and the radius of the magnetized region lies between four and five front widths out, around twice the separation of galaxies along the walls, so that the wall galaxies magnetize a continuous shell.

\begin{figure}[t]
\includegraphics[width=\columnwidth]{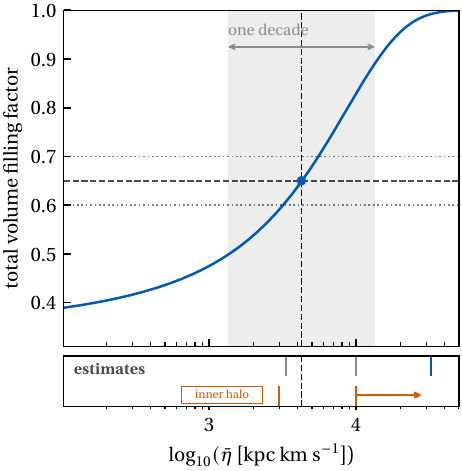}
\caption{
Total filling factor~\eqref{eq:ftotdist} as a function of $\bar\eta$ for the fiducial model. The dashed darker gray horizontal line marks the fiducial filling requirement $\freq$, and the dotted lighter gray horizontal lines its range in \cref{tab:inputs}.
The dot and the black dashed vertical line, continued through the strip, mark $\etareq$, and the shaded band spans one order of magnitude in $\bar\eta$ centered logarithmically on it.
The strip at the bottom of the plot shows various estimates of diffusivities from theory, simulations and observations. They are described in more detail in \cref{fig:requirement} and in \cref{sec:estimates}.
}
\label{fig:ftot}
\end{figure}

\Cref{fig:ftot} shows the total filling factor as a function of $\bar\eta$.
Over the decade in $\bar\eta$ centered on $\etareq$, it rises from insufficient to nearly complete, so that the magnetization of the voids responds to the void diffusivity almost as a threshold.
The steepness reflects the volume magnetized by each interior galaxy, which grows as $\rth^3 \propto \bar\eta^{3/2}$ up to the slow variation of $\Lambda$.

%-------------------------------------------------------------

\paragraph*{Contributions to the filling.}
At the required diffusivity, the shell along the walls and the interior galaxies each magnetize about a quarter of the void volume, averaged over the void sizes, and together nearly half of it (\cref{tab:fiducial}).
With the volume outside the voids, this gives $\freq$.

Without interior galaxies, $\nG \to 0$, the shell alone must cover the same fraction of the voids, and the requirement rises fourfold. This is the case we studied in \Cite{my:dipole}.
Without the shell, which fills a vanishing fraction of voids much larger than the the radius of the magnetized region, $R \to \infty$, the interior galaxies alone must do so, and the requirement nearly doubles.
Since the interior coverage~\eqref{eq:poisson} is independent of the void size, no void, however large, raises the requirement beyond this.
If the filling requirement is imposed on the voids alone, $f\sub{void} = \freq$, which amounts to $f\sub{s} = 0$, the requirement again nearly doubles.
A factor of two to four in $\bar\eta$ thus offsets the loss of an entire contribution: factors of order unity in the diffusivity matter as much as a whole population of galaxies.

\begin{figure}
\includegraphics[width=\columnwidth]{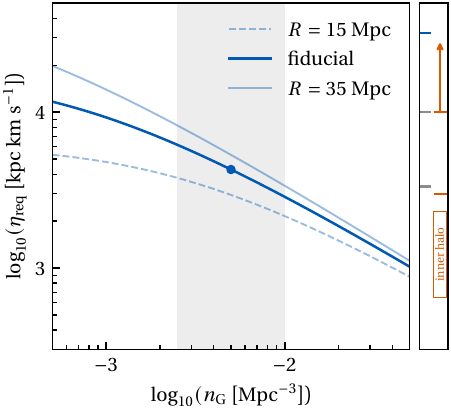}
\caption{
Minimum required diffusivity as a function of the density of galaxies inside voids, with the other inputs at their fiducial values; above the curve, the filling factor exceeds $\freq$. The shaded band marks the range of void-galaxy densities in \cref{tab:ng}, and the dot the fiducial model. The thin lines in the vertical strip on the right mark the estimates also shown in \cref{fig:requirement}, where (along with \cref{sec:estimates}) they are discussed in more detail.
}
\label{fig:density}
\end{figure}

\Cref{fig:density} shows the requirement as a function of the density of interior galaxies.
Toward low densities, it levels off at the value set by the walls alone.
Toward high densities, the interior coverage dominates; since it depends on $\nG \mk \rth^3$ with $\rth \propto \bar\eta^{1/2}$ up to the slow variation of $\Lambda$, the requirement falls as $\nG^{-2/3}$, steepened by $\Lambda$ to about $\nG^{-0.8}$.
The fiducial model lies between the two regimes, where the shell and the interior galaxies contribute comparably, so that the requirement stays within a factor of $1.5$ of its fiducial value across the range of $\nG$ in \cref{tab:inputs}.
The galaxies missing from the measured densities (\cref{sec:inputs}) would shift the fiducial point toward higher densities, where the required value of $\eta\sub{req}$ is reduced.

%-------------------------------------------------------------

Thus, within approximately half an order of magnitude from the fiducial value of $\bar \eta$, different limits of no interior galaxies at all or arbitrarily large voids or non-magnetized structure can be accommodated, underscoring the importance of the factors of order unity in the determination of the value of $\bar \eta$ for magnetizing the voids.

%-------------------------------------------------------------
\begin{figure}[t]
\includegraphics[width=\columnwidth]{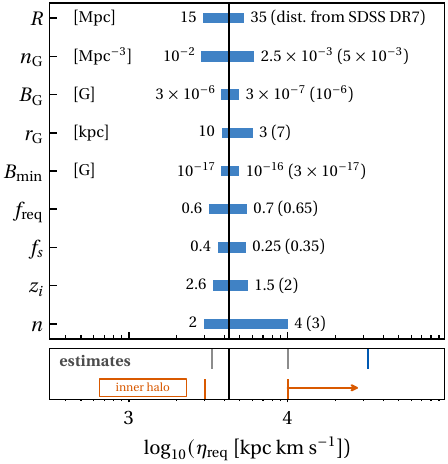}
\caption{
Minimum required diffusivity $\etareq$ for the fiducial model (vertical line) and for each input varied across its range in \cref{tab:inputs} with the others held at their fiducial values (bars). The labels at the ends of each bar give the values of the input there, with the fiducial value in parentheses; for $R$, the fiducial model uses the distribution of \cref{fig:voidsize}. The strip below shows published values of the diffusivity (\cref{sec:estimates}). Upper row: turbulent diffusivity of void gas from order-of-magnitude estimates (gray) \cite{ref:ghosh26} and from the TNG50 simulation (blue) \cite{ref:yamashita26}. Lower row (orange): diffusivity of cosmic rays in galaxy halos. The box spans the range favored for the inner halo \cite{tex:crocker26}. The two lines mark, from left to right, the value at the edge of the halo \cite{ref:yamasaki24} and the lower limit in the outer halo, drawn at the middle of its range (arrow) \cite{ref:butsky23}. More details can be found in \cref{sec:estimates}.
}
\label{fig:requirement}
\end{figure}

%-------------------------------------------------------------
\subsection{Range of the requirement}
\label{sec:req_range}

\Cref{fig:requirement} shows the required diffusivity when each input is moved across its range with the others held at their fiducial values.
The void size $R$, the galaxy density $\nG$, the volume fraction outside voids $f\sub{s}$, and the filling requirement $\freq$ do not enter the reach, and each keeps the requirement within a factor of $1.5$ of its fiducial value.
Since the walls and filaments count as magnetized, a smaller $f\sub{s}$ leaves more of the volume to be magnetized by galaxies and raises the requirement.
The start redshift $z_i$ enters only through the diffusion time~\eqref{eq:dsfactor}, so that $\etareq \propto 1/q\sub{H}(z_i)$ exactly, which stays within a factor of $1.3$.
The galactic field, characterized by $B\sub{G}$ and $r\sub{G}$, and the lower bound $\Bmin$ enter only logarithmically, through $\Lambda$ in \cref{eq:rth}, so each stays within a factor of $1.2$.
The field geometry has the largest effect.
It changes the power $n$ of $r\sub{G}/\rs \approx 10^{-2}$ in $\Lambda$, shifting $\Lambda$ by about four from its fiducial value of about ten; the requirement rises by a factor of $2.3$ for the poloidal quadrupole and falls by about a third for the toroidal component.

\paragraph*{Combined inputs.}
Moving the field geometry, the galaxy density, and the start redshift together to their favorable ends lowers the requirement to about $\num{1.6}\,\mpckms$, and to their adverse ends raises it to about $\SI{2.5e1}{\mpckms}$.
With all inputs at the ends of their ranges and the dipole geometry, the requirement lies between about $\num{0.7}$ and $\num{7e1}\,\mpckms$, approximately within an order of magnitude of its fiducial value.
Therefore, whether galaxies magnetize the voids rests mainly on the void diffusivity with $\order{1}$ factors fully determined.

%-------------------------------------------------------------
\subsection{Comparison with estimates and observations}
\label{sec:estimates}

Spitzer resistivity gives diffusion lengths of order $10^{-4}\,$pc \cite{ref:seller26}, for which galactic fields remain confined to their galaxies, so that the magnetization of voids by galaxies rests entirely on turbulent transport.
In astrophysical plasmas, turbulence transports large-scale magnetic fields at rates that exceed the microphysical ones by many orders of magnitude, $\eta \simeq \eta\sub{t}$ (\cref{sec:eta}).
Estimates of this form for void gas, \SI{3.3}{\mpckms} and \SI{10}{\mpckms} \cite{ref:ghosh26}, are marked as gray lines in the strip of \cref{fig:density,fig:ftot,fig:requirement}: the fiducial requirement lies between them, and the required $\bar\eta$, with single inputs varied across their ranges, spans the interval.
A redshift-dependent $\eta\sub{t}$ has also been obtained from the turbulent velocities and scales of the intergalactic gas in the TNG50 simulation \cite{ref:yamashita26}; the constant diffusivity that accumulates the same diffusion time~\eqref{eq:clock} for $z_i = 2$, \SI{3.2e1}{\mpckms}, exceeds the requirement even with any of the inputs at their adverse ends (\cref{sec:req_range}). This diffusivity is shown as the blue line in the strips of \cref{fig:density,fig:ftot,fig:requirement}.

In the voids themselves, no observation constrains the turbulent diffusivity.
In galaxy halos, observations constrain a related quantity, the diffusivity $\kappa\sub{CR}$ of cosmic rays, also marked in the lower half of the strip.
Cosmic rays are tied to magnetic field lines: they move along the lines, and they move with the lines when turbulence displaces them \cite{ref:krumholz20,ref:sampson23}.
In halo gas, the motion along the lines, at the Alfv\'en speed, is slower than the turbulent motion of the gas \cite{tex:crocker26}.
The turbulence therefore carries the cosmic rays as it carries the field, so that $\kappa\sub{CR}$ measures $\eta\sub{t}$ up to factors of order unity.
For a halo of Milky Way mass, subsonic turbulence with velocities of about \SI{30}{km.s^{-1}} \cite{tex:crocker26} and eddies of about \SI{100}{kpc} gives $\eta\sub{t} \sim \SI{1}{\mpckms}$.
This is three to four orders of magnitude above the value in galactic disks, where the velocities are comparable and the eddies are about a hundred parsecs across \cite{ref:brandenburg23}. The diffusivity of cosmic rays is likewise found to rise with distance from the galaxy \cite{ref:butsky23}.

For the halo of a Milky Way-mass galaxy as a whole, $\gamma$-ray observations and the pressure balance of the gas favor $\kappa\sub{CR}$ between \SI{6.5e-1}{\mpckms} and \SI{2.3}{\mpckms} \cite{tex:crocker26}, in agreement with the value of $\eta\sub{t}$ estimated above.
The observations that constrain it probe mostly the dense inner halo. This range is shown in the strips of \cref{fig:density,fig:ftot,fig:requirement} as a box denoted by the text ``inner halo."
In the outer halo, absorption-line measurements give lower limits of \SI{3}{\mpckms} to \SI{3e1}{\mpckms} \cite{ref:butsky23}. This range of the lower limit is denoted by a single value (\SI{e1}{\mpckms}) to represent the lower limit; in the strips of \cref{fig:density,fig:ftot,fig:requirement}, it is shown as  the rightmost orange line (with an arrow).
The transport thus becomes faster outward, as the constraints on the halo-wide value allow \cite{tex:crocker26}.
At the edge of the halo of more massive galaxies, the radio shells of odd radio circles, interpreted as virial shocks, give $\kappa\sub{CR} \sim \SI{3}{\mpckms}$ \cite{ref:yamasaki24}.
In the halo, the relation $\eta\sub{t} \simeq v \mk \ell / 3$ thus reproduces the observed diffusivity. This observation is shown as the left most orange line in the strips of \cref{fig:density,fig:ftot,fig:requirement}.
This supports the estimates for void gas, which rest on the same relation and lie within an order of magnitude of all the values observed in halos.

The fiducial requirement lies inside the range spanned by these values.
The estimates of \Cite{ref:ghosh26}, the lower end of the outer-halo limits, and the value at the edge of the halo all lie within a factor of about two of it.
Only the halo-wide value, weighted toward the inner halo, lies as much as a factor of seven below it, and only the upper end of the outer-halo limits and the estimate of \Cite{ref:yamashita26} as much as a factor of eight above.
Beyond the halo, where the virial radius no longer limits the size of the eddies, void gas thus needs no larger a diffusivity than is found in the outer halo.
Since the filling rises from insufficient to nearly complete within the range of these values (\cref{sec:req_fiducial}), settling whether galaxies magnetize the voids requires determining the effective diffusivity of void gas including its $\order{1}$ factors.

%%%%%%%%%%%%%%%%%%%%%%%%%%%%%%%%%%%%%%%%%%%%%%%%%%%%%%%%%%%%%%
\section{Discussion}
\label{sec:discussion}

The required diffusivity calculated here rests on assumptions that are mostly conservative, in that relaxing them lowers it.
First, we neglect the superposition of fields from neighboring galaxies, which raises the RMS field wherever they overlap.
Second, we hold the comoving dipole moment at its present value, whereas a bound galaxy of constant physical moment carried a larger comoving moment at earlier times.
Third, we model the voids as spheres, which at fixed volume have the least volume within a given distance of the boundary; irregular voids receive a larger coverage from their walls than \cref{eq:shell} implies.
Fourth, our value of $f\sub{s}$ assigns to the voids more of the volume than the catalogs underlying the galaxy densities, and we populate the excess, which lies at the denser void edges, at the interior density (\cref{sec:inputs}).
Finally, we include only diffusion, whereas galactic winds and outflows from active galactic nuclei also carry magnetized gas beyond the halos \cite{ref:bertone06,ref:samui18,ref:aramburogarcia21,ref:hopkins21} and can only add to the magnetized volume.
On the other hand, the clustering of void galaxies raises the requirement: they trace tenuous filaments \cite{ref:alpaslan14}, so their magnetized regions overlap and the interior coverage~\eqref{eq:poisson} falls at fixed density.

The blazar bound is weakest for fields coherent over more than $\SI{0.2}{Mpc}$ and strengthens for shorter coherence lengths \cite{ref:acciari23}.
Beyond the diffusion front, the field of a single galaxy is antiparallel to the component of its dipole moment perpendicular to the radial direction~\eqref{eq:farfield}.
Its direction is constant along each radial line and varies only with the angular position around the galaxy.
The field is therefore coherent over a length comparable to the distance from the galaxy, which is a few megaparsecs at the surface of the magnetized region.
Across the void, the direction changes between the regions magnetized by different galaxies, on the scale of their separation.
In both cases the coherence length exceeds $\SI{0.2}{Mpc}$ by a wide margin.

Fields of this strength lie far beyond the reach of radio observations.
Faraday rotation of background radio sources limits the field in voids to the nanogauss level \cite{ref:heald20,ref:mtchedlidze24}, seven orders of magnitude above the blazar bound.
The rotation-measure grid of the Square Kilometre Array will increase the density of sightlines a hundredfold \cite{ref:heald20}, which sharpens these limits but without closing the gap.
The spatial structure of the void field, which would reveal its origin, thus cannot be mapped, and galaxy-seeded void fields must be identified through other properties that distinguish them from a primordial field.

One such property is the growth of the void field with cosmic time.
Since the field spreads by diffusion, a void observed at higher redshift has had less diffusion time since the galactic fields formed~\eqref{eq:clock}, and the radius of the magnetized region and filling factor are correspondingly smaller.
Fields generated in the early Universe, in contrast, retain an approximately constant comoving strength on megaparsec scales after recombination \cite{ref:durrer13}.
A galaxy-seeded field therefore weakens toward higher redshift, while a primordial field does not. The Cherenkov Telescope Array will search for the extended $\gamma$-ray halos that the void field forms around blazars by deflecting the cascade pairs \cite{ref:abdalla21}, and halos measured over a sufficiently wide range of redshift could distinguish the two origins.

A second is the dependence on void size.
The coverage from the walls decreases with the size of a void, and in the largest voids the filling approaches that of the interior galaxies alone (\cref{sec:coverage}).
The fraction of a void that remains below the bound therefore grows with its size and lies at its center.
A primordial field fills voids of all sizes alike, so that a dependence of the measured field on the sizes of the voids along the line of sight would indicate a galactic origin.

The two origins are not exclusive, since the contributions of a galaxy-seeded and a primordial field add in the voids.
A primordial field then need not supply the entire volume-filling field implied by the blazar observations, which relaxes the requirements on its generation in the early Universe.
The galaxy-seeded field, in turn, would act as a foreground to the primordial one.
As the foregrounds of the cosmic microwave background are separated from the primordial signal through their dependence on frequency, the galactic contribution can be separated through its dependence on redshift and on void size, and it has to be modeled before the void field is used to constrain the early Universe.

%%%%%%%%%%%%%%%%%%%%%%%%%%%%%%%%%%%%%%%%%%%%%%%%%%%%%%%%%%%%%%
\section{Conclusions}
\label{sec:conclusions}

We have determined the effective magnetic diffusivity of void gas required for galaxies to magnetize the voids to the level inferred from the $\gamma$-ray cascades of TeV blazars.
In a turbulent gas, galactic fields spread outward behind a Gaussian front whose width is set by the diffusion time accumulated since the fields formed, including the effect of cosmic expansion.
The radius of the magnetized region scales linearly with this width and only logarithmically with the galactic field strength, the observational bound, and the field geometry~\eqref{eq:rth}.

The required diffusivity is about \SI{4.3}{\mpckms} for fiducial inputs based on observations of void galaxies, void sizes, and galactic fields (\cref{tab:inputs,tab:fiducial}).
Galaxies on the walls of a void and those in its interior magnetize complementary regions and contribute comparably at the required diffusivity.
In large voids, the requirement approaches a finite value set by the interior galaxies alone.
The filling factor rises from insufficient to nearly complete over one decade in the diffusivity (\cref{fig:ftot}), so that a factor of about~2--4 offsets the loss of either population.
Each input, varied across its plausible range, changes the requirement by less than a factor of~$1.5$, except the field geometry, which changes it by a factor of~2.3 (\cref{fig:requirement}).

The requirement lies inside the range spanned by the estimates for void gas and by the cosmic-ray transport observed in galaxy halos, most of which lie within a factor of 2 of it (\cref{fig:requirement}).
In the halos, the turbulent diffusivity estimated from the velocities and sizes of the eddies reproduces the observed transport, which rises with distance from the galaxy.
In the outer halo, the outermost gas probed, the observations give lower limits, and these already reach the requirement.
If the diffusivity of void gas is compatible with these limits, galaxies magnetize the voids to the level inferred from the blazar observations.
The effective diffusivity of void gas, determined to within factors of order unity, then fixes how much of the void field is galactic.

The blazar observations constrain the total field in the voids, the sum of a galactic and a primordial contribution.
A primordial field thus need not supply the entire volume-filling field, which relaxes the requirements on its generation in the early Universe.
The galactic contribution acts as a foreground to the primordial one, distinguished by its growth with cosmic time and its decline with the size of the void.
Across the diffusivities estimated for void gas and observed in halos, this foreground ranges from a part of the volume-filling field to nearly all of it.
Whether the blazar observations require a primordial field at all therefore depends on the diffusivity of void gas.

%%%%%%%%%%%%%%%%%%%%%%%%%%%%%%%%%%%%%%%%%%%%%%%%%%%%%%%%%%%%%%
\section*{Acknowledgements}

This study was motivated by discussions with Mohammadreza Ayromlou, Axel Brandenburg, Chiara Caprini, Andrii Neronov and Kandaswamy Subramanian.
DL and JS  acknowledge funds by the Deutsche Forschungsgemeinschaft (DFG, German Research Foundation) under Germany’s Excellence Strategy – EXC 3037 – 533607693.


\end{document}